%% file: main.tex
\documentclass[letterpaper,twocolumn,10pt]{article}
\usepackage{usenix}

\usepackage{balance}
  
\usepackage[utf8]{inputenc}
\usepackage{listings}
\usepackage{graphicx}
\usepackage{svg}
\usepackage[frozencache,cachedir=minted-cache]{minted}
\usepackage{amssymb}
\usepackage{pifont}
\usepackage{enumitem}
\usepackage{tabularx} 
\usepackage{adjustbox}
\usepackage{tikz}
\usepackage{pgfplots}
\usepackage{appendix}
\usepackage[caption=false]{subfig}
\usepackage{threeparttable}
\usepackage{makecell}
\usepackage{multirow}
\usepackage{algorithm}
\usepackage{algorithmicx}  
\usepackage[noend]{algpseudocode}
\usepackage{appendix}
\usepackage{hyperref}
\usepackage{breakurl}

\graphicspath{ {./images/} }

\newcommand{\toolname}{\code{Patch2QL}}
\newcommand{\qltoolname}{CodeQL}
\newcommand{\qlname}{\code{QL}}

\newcommand{\ignore}[1]{}

\newcommand{\tab}{\hspace*{1em}}
\newcommand{\code}[1]{{\fontfamily{cmtt}\fontseries{m}\fontshape{n}\selectfont\small{#1}}}

\begin{document}

\title{\Large \bf Patch2QL: Discover Cognate Defects in Open Source Software Supply Chain With Auto-generated Static Analysis Rules}

\author{
{\rm Fuwei Wang\thanks{Corresponding author. Email: \href{mailto:fullwaywang@outlook.com}{fullwaywang@outlook.com}}, Yongzhi Liu, Zhiqiang Dong}\\
Yunding Lab, Tencent Inc.
}

\maketitle

\input{abstract.tex}

\input{introduction.tex}

\input{background.tex}

\input{architecture.tex}

\input{keyTechniques.tex}

\input{evaluation.tex}

\input{discussion.tex}

\input{conclusions.tex}

\bibliographystyle{plain}
\bibliography{main}

\input{appendix}

\end{document}

%% file: abstract.tex
\begin{abstract}

In the open source software (OSS) ecosystem, there exists a complex software supply chain, where developers upstream and downstream widely borrow and reuse code. This results in the widespread occurrence of recurring defects, missing fixes, and propagation issues. These are collectively referred to as cognate defects, and their scale and threats have not received extensive attention and systematic research. Software composition analysis and code clone detection methods are unable to cover the various variant issues in the supply chain scenario, while code static analysis, or static application security testing (SAST) techniques struggle to target specific defects.

In this paper, we propose a novel technique for detecting cognate defects in OSS through the automatic generation of SAST rules. Specifically, it extracts key syntax and semantic information from pre- and post-patch versions of code through structural comparison and control flow to data flow analysis, and generates rules that matches these key elements. We have implemented a prototype tool called \toolname{} and applied it to fundamental OSS in C/C++. In experiments, we discovered $7$ new vulnerabilities with medium to critical severity in the most popular upstream software, as well as numerous potential security issues. When analyzing downstream projects in the supply chain, we found a significant number of representative cognate defects, clarifying the threat posed by this issue. Additionally, compared to general-purpose SAST and signature-based mechanisms, the generated rules perform better at discover all variants of cognate defects.

\end{abstract}

%% file: introduction.tex
\section{Introduction}
\label{sec:introduction}

Open source software (OSS) has been the cornerstone of the entire Internet ecosystem. The open-source code from upstream developers influences the downstream ecosystem in various forms, such as being invoked, integrated, copied and partially reused, forming a complex software supply chain. According to a recent report by Synopsys~\cite{ossra}, $96\%$ of scanned codebases across 17 industries were found to contain open source components. With widespread usage and scrutiny, the OSS ecosystem has witnessed an increase in emerging vulnerabilities and even more potential code bugs that could pose security issues, which we collectively refer to as \textit{defects}. The issue of software defects has further given rise to new problems in the software supply chain context. One of them is the emergence of cognate defects, which is accompanied by various practices in the software supply chain, such as open source code reuse, replicating code with similar functionality, project cloning and secondary development, fragmented software distribution and maintenance, and so on.

The goal of discovering cognate defects can be divided into several missions: a) For an upstream OSS, identifying recurring defects that are similar to the original ones, indicating the presence of the same coding mistakes; b) For a forked project based on an OSS, confirming whether upstream defects exist and affect the current codebases, as well as determining if any newly added modules that borrow code from the original codebases contain derived defects; and c) For any project, verifying if it contains similar defects to the known ones in OSS, either in syntactic or semantic senses.

However, finding cognate defects is not as straightforward as it may appear. The effectiveness of Software Component Analysis (SCA) mechanisms heavily relies on explicit dependency analysis, which is not applicable in C/C++ ecosystems. In the research field, the discovery of cognate defects is classified as a code clone detection problem, and there has been a significant amount of related research work in recent years. The mainstream idea is to use customized code fingerprints or signatures to represent defect characteristics and to design corresponding fuzzy matching algorithms to identify target code with textual or semantic similarities. However, as we will discuss below, syntactic similarity only reveals a limited portion of simple similar defects, and the use of similarity-based methods is greatly restricted by their high demand for computing power. Static Application Security Testing (SAST) are widely used to match specific code patterns, effectively identifying defects with the same root cause. Nevertheless, the significant gap between Common Vulnerabilities and Exposures (CVEs) and available SAST rules should not be overlooked, as most SAST tools focus on detecting common types of weaknesses, while a significant portion of real-world vulnerabilities result from a combination of common principles and specific logic.

\smallskip \noindent
\textbf{Our approach}\tab
In this paper, we propose a method to discover \textit{cognate defects} in OSS. The underlying concept is that the root cause of a defect can be inferred from the differences between pre- and post-patched code. The main idea is to understand the root cause of a defect by analyzing its patch, and then automatically generate a SAST rule other than some kind of signatures, which can specifically match the unpatched target code and its control-flow and data-flow context. This capability allows us to identify code that are syntactically or semantically similar.

First, we collect historical vulnerabilities and corresponding patches. We retrieve the Abstract Syntax Trees (ASTs) of both the unpatched and patched codes to be compared, taking into account the hierarchically structured information. Secondly, we filter and reorganizes all the differing nodes, transforming them into a raw SAST rule, which matches all occurrences in the code where standalone unpatched nodes and their contexts exist, while new nodes introduced in the patch do not. Then, our system establishes the control- and data-flow context of the differing AST nodes by traversing the shared sections of the two ASTs around those nodes. In terms of control-flow context, we identify the structurally relevant nodes that indicate the position of a differing node. For the data-flow context, we locate the important data variables referenced in the differing node and perform a straightforward data-flow analysis around them. Finally, through regression tests, we refine the rule by simplifying and enriching the set of queries and conditions in order to effectively match the unpatched code. 

\smallskip \noindent
\textbf{Prototype and evaluation}\tab
We have developed a prototype tool \toolname{} specifically targeting OSS in C/C++. For the underlying SAST technique, we have chosen the promising tool CodeQL~\cite{codeql} and its query-like language \qlname{}~\cite{de2007keynote}. We selected $111$ projects in C language from the Google OSS-Fuzz project's list~\cite{ossfuzzprojects} as our targets and built a repository of generated \qlname{} rules. We evaluated the effectiveness of the rules against 8 most popular projects, and results show that the recall rate of these rules can reach $71\%$, and the detection accuracy is significantly higher than that of the compared SAST tool and signature-based code clone detection methods.

We conducted an investigation into the OSS ecosystem to uncover the prevalence and severity of cognate defects in the OSS supply chain. We applied our rules to two types of targets: As for \textit{upstream} OSS projects, we have identified $7$ recurring new CVEs cognate to existing ones. Some other issues found were reported and got fixed. As for \textit{downstream} OSS projects, we analyzed popular open-source database projects (including forks or projects suspected of borrowing code from upstream OSS) and three Linux distribution kernels primarily developed and maintained by cloud computing vendors. We detected a batch of known defects or their variants that remain unfixed in these projects, some of which exhibit little contextual similarity with their source.


In summary, this paper makes the following contributions.

\begin{itemize} [leftmargin=*]
    \item We conducted a comprehensive study of various types of cognate defect types of different levels and causes, and explained the current state of the problem through the discovery of corresponding new vulnerabilities.
    
    \item We innovatively use automatically generated SAST rules as features to detect cognate defects, which provides unique detection capabilities compared to general-purpose SAST rule-sets and code clone detection mechanisms based on custom signatures.
    
    \item We have developed and implemented a prototype specifically designed for C language. Over the course of less than two months, we applied this prototype to analyze several fundamental OSS projects. During this period, we successfully identified and reported $7$ zero-day vulnerabilities, in addition to numerous other issues.
\end{itemize}

The rest of this paper is structured as follows: we introduce the background and the defined threat model in Section~\ref{sec:background}, and illustrate the overall design of our system in Section~\ref{sec:architecture}. We further illustrate the key techniques and implementation of our system in Section~\ref{sec:keytech}. We then present the evaluation result in Section~\ref{sec:evaluation}. At last, we discuss the limitations in Section~\ref{sec:discussion} and conclude our work in Section~\ref{sec:conclusions}.

%% file: background.tex
\section{Scope of the Problem}
\label{sec:background}

\subsection{Roles in OSS supply chain}
The roles of suppliers and consumers can be categorized into 3 layers. An \textit{upstream} supplier is the source of OSS codes. Developers code and maintain the sources, resolve issues, and merge contributions from individuals. The \textit{midstream} layer involves secondary developers who fork and redistribute diverged copies. There are also platforms where OSS are manufactured to be distributed to downstream users. Depending on how OSS is used, there are three types of \textit{downstream} consumers: those who directly use OSS as binaries and SDKs; those who integrate OSS as intact components into their programs explicitly; and those who borrow pieces of OSS source code (modules, files or snippets) into their own projects.

Several types of threats around specific roles have been well studied and addressed. For instance, SCA and SBOM are mature techniques for identifying and recording intact components a downstream project imported or linked. Cognate defects, on the other hand, have been rarely studied systematically or addressed adequately, which we try to explain in the following.


\subsection{Cognate defects threat model}
\label{subsec:threatmodel}

We define \textit{cognate defects} as defects that have the same root cause. In certain contexts, cognate defects are referred to by different terms. Bugs that propagate through software reuse are commonly known as \textit{recurring} bugs~\cite{pham2010detection, kang2022tracer}, indicating that known and previously fixed bugs resurfaced in other codebases. Some are referred to as \textit{variants of known vulnerabilities}~\cite{zerodayexp}, which include partially fixed vulnerabilities, regressions in patched code, or similar bugs found in different subsystems of Linux.

Depending on the forms and degrees of similarity between \textit{cognate} defects in source code, we provide a four-level classification:

\begin{description} [leftmargin=*]
    \item[L1] The vulnerable codes are almost the same except for minor textual modifications, while relevant code context (files, modules and functions) may contain minor changes, such as the changes introduced between two near versions without significant development.
    \item[L2] The vulnerable codes remain intact (sometimes irrelevant codes can be inserted in the middle), but there are major changes in the context, such as those modifications brought about by code refactoring.
    \item[L3] Only the defect-related target code fragments are reused, and the code fragments and their contexts may have been rewritten or refactored; or the defect is composed of a specific pattern of statement sequences dominated by function calls.
    \item[L4] Defects are only reused or shared at the level of functional logic and semantics, primarily including the presence of the same defects in code implementations that utilize the same algorithm or protocol standards.
\end{description}

Through an extensive analysis of various cases, we have summarized the representative sources and means of propagation of cognate defects in the development process of all roles of OSS supply chain:

\smallskip \noindent
\textbf{False regression}\tab
Bug patches can be mistakenly removed through collaboration or code reconstruction due to the lack of bug tracking. A fixing commit pushed into a development branch can also be occasionally reverted by a conflicting commit. These defects, if found in time, fall into L1--L2 type. Such cases had been extensively discussed in \cite{zerodayexp}.

\smallskip \noindent
\textbf{Fragmentation}\tab
Users of OSS typically adopt a snapshot version and often engage in custom development based on specific needs. The divergence compared to the upstream development branch increases over time, making it challenging to keep them in sync. This situation makes it difficult to port patches. These unfixed defects fall into L2 types. The example we discuss in detail in Section~\ref{subsec:kernel} best illustrates this scenario.

\smallskip \noindent
\textbf{Copycat}\tab
For collaborative or forked projects, subsequent developers may develop new modules with similar logic based on existing code, resulting in the inheritance of defects in the process. These defects fall into L2--L3 types. Particularly, the seeming bugs mentioned in Table~\ref{tab:forksdb} can be classified into this category.

\smallskip \noindent
\textbf{Common weakness}\tab
In implementations of similar logic, common weakness types can lead to similar or repeated defects. Additionally, unclear interface documentation can mislead developers and cause them to misuse the API. These mostly fall into L3--L4 types. The 2 instances cognate to CVE-2020-29599 we discovered illustrates this scenario, as described in Table~\ref{tab:cvelist}.

\smallskip \noindent
\textbf{Imitated poisoning}\tab
Though not explicitly discussed, it is realistic to poison OSS by introducing a crafted code snippet that is similar to a previously vulnerable one. It is challenging for maintainers to identify if a commit, appearing at a later time, is an imitation of a previous vulnerability. These rare cases might fall into L4 types.

\subsection{Burden of OSS in C/C++}
\label{sec:burden}
Considering the above analysis, cognate defects are more prominent in development language ecosystems with longer histories, as their OSS supply chains are more complex and opaque. Newer programming languages, such as Golang, have more centralized support for basic libraries/packages and have not yet widely adopted code replication practices. In ecosystems like Java, which has centralized package management like Maven and authoritative foundations like Apache that provide a large number of commonly used components, composition dependency is relatively clear. However, in the case of C/C++, which has a long and foundational history with many coexisting projects with similar functionalities and lacks a general package management mechanism, cognate defect issues have become a huge burden. 

The dominance of C/C++ in foundational systems and applications remains unshakeable, and the problem models it faces may exist in other ecosystems to a lesser extent or in near future. Therefore, our research focuses on the hardcore ecosystem of C/C++ and ended up with the implementation of prototype tools for C-language OSS to address C/C++ issues. However, the methodological approach and the prototype tool itself can be adapted for discovering potential problems in other language ecosystems after certain adjustments.

%% file: architecture.tex
\section{System Overview}
\label{sec:architecture}

\subsection{Motivation}
The relationship between the root cause, immediate cause, and patch of a vulnerability has been extensively studied for many years. Previous research~\cite{zhao2020patchscope} has observed that, for common vulnerability types such as memory corruption, patches reveal significant differences in how input data is manipulated between different versions, providing valuable semantic information.

The motivation behind our work stems from the understanding that a patch, along with its control-flow and data-flow context, contains information about the root cause of a defect. By abstracting and associating this information, we can uncover the underlying semantic causes implied by syntactical differences. Given the complexity of software, we believe that the types and portions of patches carrying sufficient implications of root causes cannot be theoretically enumerated and summarized. However, even for vulnerabilities where root causes cannot be inferred solely through fixes, patches can still be utilized to identify syntactically similar code segments, which may overlap with actual cognate defects.

\begin{listing}[tb]
    \begin{minted}[frame=lines, breaklines,
                       fontsize=\scriptsize,
                       tabsize=2]{diff}
diff --git a/crypto/x509/v3_utl.c b/crypto/x509/v3_utl.c
index 4fd1f2cd60..5c63d2d9d8 100644
--- a/crypto/x509/v3_utl.c
+++ b/crypto/x509/v3_utl.c
@@ -529,17 +529,25 @@ static int append_ia5(STACK_OF(OPENSSL_STRING) **sk,
     /* First some sanity checks */
     if (email->type != V_ASN1_IA5STRING)
         return 1;
-    if (!email->data || !email->length)
+    if (email->data == NULL || email->length == 0)
+        return 1;
+    if (memchr(email->data, 0, email->length) != NULL)
         return 1;
     if (*sk == NULL)
         *sk = sk_OPENSSL_STRING_new(sk_strcmp);
     if (*sk == NULL)
         return 0;
+
+    emtmp = OPENSSL_strndup((char *)email->data, email->length);
+    if (emtmp == NULL)
+        return 0;
+
     /* Don't add duplicates */
-    if (sk_OPENSSL_STRING_find(*sk, (char *)email->data) != -1)
+    if (sk_OPENSSL_STRING_find(*sk, emtmp) != -1) {
+        OPENSSL_free(emtmp);
         return 1;
-    emtmp = OPENSSL_strdup((char *)email->data);
-    if (emtmp == NULL || !sk_OPENSSL_STRING_push(*sk, emtmp)) {
+    }
+    if (!sk_OPENSSL_STRING_push(*sk, emtmp)) {
         OPENSSL_free(emtmp); /* free on push failure */
         X509_email_free(*sk);
         *sk = NULL;
    \end{minted}
    \caption{The patch of CVE-2021-3712 in OpenSSL.} 
    \label{list:opensslpatch}
    \vspace{-0.2em}
\end{listing}

\smallskip \noindent
\textbf{Running example}\tab
In the following introduction of our algorithm and system, we use a running example for a clear illustration. The example is taken from a historical vulnerability CVE-2021-3712~\cite{cve20213712} of OpenSSL. This is a buffer overrun vulnerability caused by falsely regarding the buffer within an ASN.1 strings as \code{NULL} terminated. A patch fixing one occurrence of the bug in function \code{append\_ia5} is commit \code{986247}, as shown in Listing~\ref{list:opensslpatch}.

\subsection{Architecture}

\begin{figure*}[t]
    \centering
    \includegraphics[width=0.95\linewidth]{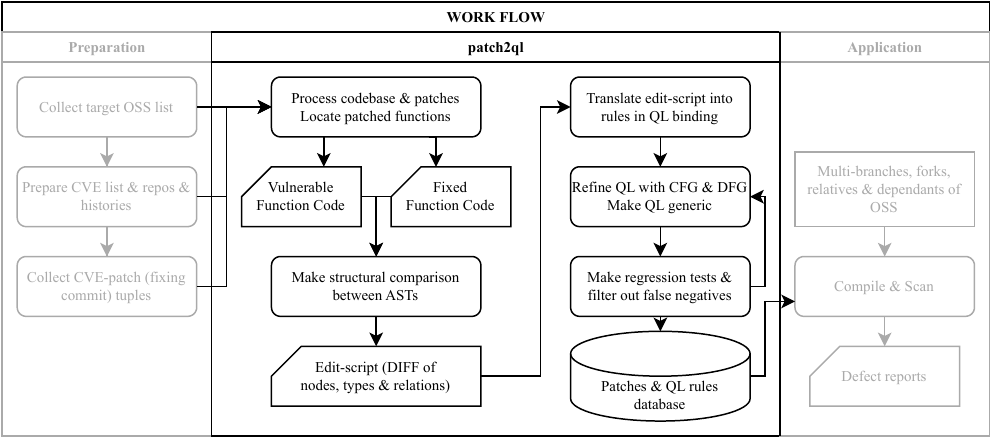}
    \caption{The core actions, prerequisites and application of \toolname{}.}
    \label{fig:architecture}
\end{figure*}

The architecture of \toolname{} is illustrated in Figure~\ref{fig:architecture}. Essentially, \toolname{} is a tool that generates SAST rules based on patches that fix security defects. Its main task is to \textit{transform} various elements and relationships between vulnerable and patched function pairs into rules. These rules can then be used to match potentially vulnerable functions.

Currently, we utilize \qltoolname{} as the underlying SAST tool and rule format binding. As a SQL-like query language, \qltoolname{} offers great expressiveness and is developer-friendly. Constructing, selecting, and organizing queries with conditions forms the basis of creating sound rules. However, the core techniques of \toolname{} are not limited to \qltoolname{} alone. If provided with another SAST tool along with its rule formats, generating rules in its specific binding is straightforward.

When given a project and a security patch, \toolname{} follows five main procedures to generate a rule, which are detailed as follows.

\subsubsection{Process the patch and locate function pairs}
The representation of a vulnerability is performed at the function level. To capture the differences between a function before and after a patch being applied in terms of basic syntax units, the codebases need to be transformed into abstract syntax trees (ASTs). Additionally, a patch typically consists of a textual listing detailing the modifications made to the vulnerable codebase. We locate the precise functions being modified by traversing and comparing all the ASTs of functions within the patched source code file.

Furthermore, there are instances when patches require preprocessing before being processed further. Some OSS maintainers often submit large code commits that address multiple bug fixes. In other cases, a patch for a single vulnerability may fix multiple instances of the same root cause. In such scenarios, we extract the minimal changes necessary to fix each individual bug. Currently, this is mainly accomplished through simple text redundancy analysis. It involves analyzing whether a patch contains multiple instances of repeated fixing patterns and splitting them into separate patches. For example, if a patch fixes multiple occurrences of the same API misuse issues, each fix of callsite will be separated. Additionally, certain vulnerabilities may require several commits to be fully resolved, necessitating the merging of these changes to produce a complete patch.

\subsubsection{Make structural comparison between ASTs}
To obtain an accurate list of differences between the ASTs of the pre- and post-patched function, a structural comparison is required. In our implementation, ASTs are generated by \qltoolname{} through compilation. Expressions, statements, literals, function calls are named \textit{AST nodes}. Taken the sub-trees of the \code{if-then} branch in Lines \code{@@ -539,2 +546,4 @@} of the patch shown in Listing~\ref{list:opensslpatch} as an example, the difference appears to be replacement of the the second parameter node of function call. An illustration of ASTs can be referred to as Figure~\ref{fig:asts} in the appendix.

The task of comparing ASTs is accomplished using the EditScript algorithm~\cite{chawathe1996change}, which is widely utilized for comparing structured information. Specifically, we propose a customized method within the algorithm to find a matching between two ASTs, which is further explained in detail in Section~\ref{sec:editscript}. After that, we obtain an editscript that reveals the necessary edit operations to transform the pre-patched AST into the post-patched AST. These \textit{operations} encompass four basic types: insertions, deletions, updates, and moves. In the \code{if-then} branch mentioned above, the node \code{`[VariableAccess] emtmp`} is deleted from the sub-tree of the \code{`call to ossl\_check\_OPENSSL\_STRING\_type`}, while a new sub-tree representing \code{`email->data`} is inserted. A wrapper node of type \code{BlockStmt} is inserted, and the original sub-tree of the \code{return} statement is moved here as the right-cousin of the newly inserted sub-tree of the \code{`call to CRYPTO\_free`}. Notably, we handle two types of moves: moving a node to become a child of a different parent node and changing the order under the same parent node.

\subsubsection{Transform editscript into rules}
To produce a raw rule in the \qlname{} binding from an editscript, we need to construct a query that targets each differential AST node. This query should include specific conditions that restrict the attributes and position of the node to match a corresponding defect.

We utilize a generic template and a subset of the \qlname{} grammar as the protocol for generating rules. Essentially, the queries for nodes that correspond to the four types of edit operations can be expressed as the \textit{existence} or \textit{nonexistence} of targets. In other words, we need to match a function where the inserted elements do not exist and the deleted elements are present, while the updated and moved nodes have the same attributes and position as they do in the unpatched AST. To achieve this, we define a \textit{predicate} for each differential node, which is a set of query conditions presented in a form similar to a function.

\begin{listing}[tb]
    \begin{minted}[frame=lines,
                    breaklines,
                    breakafter=.(=,
                    fontsize=\scriptsize,
                    tabsize=2]{sql}
predicate func_0(Parameter vemail_525, FunctionCall target_0) {
	target_0.getTarget().hasName("CRYPTO_strdup")
	and not target_0.getTarget().hasName("CRYPTO_strndup")
	and target_0.getArgument(0).(PointerFieldAccess).getTarget().getName()="data"
	and target_0.getArgument(0).(PointerFieldAccess).getQualifier().(VariableAccess).getTarget()=vemail_525
	and target_0.getArgument(1) instanceof StringLiteral
	and target_0.getArgument(2) instanceof Literal
}
    \end{minted}
    \caption{An example raw \qlname{} predicate.} 
    \label{list:predicate}
    \vspace{-0.2em}
\end{listing}

To determine the conditions that a node must satisfy, it is important to reference data objects and describe the relationships between them and the target. This association relationship is memory objects (variables) - centric. In other words, it defines all the necessary variable objects and describes the relationships between all AST nodes and associated variables. These key variables are identified throughout the entire rule by their original variable names together with declaration locations, and only their types are used as query qualifiers, ensuring the ability to distinguish the target variable objects without being affected by variable renaming or aliases. By considering the relationships among target nodes and variables, we can outline the relative positions. The path from a node to a related variable can be extracted from the AST. For example, a predicate can be constructed to describe an updated node in a function call that replaces the function \code{CRYPTO\_strndup} with \code{CRYPTO\_strdup}, as shown in Listing~\ref{list:predicate}.

\subsubsection{Refine \& generalize raw \qlname{} rules}
A raw rule is often not accurate enough to precisely target a specific node, which can result in both over-matching and under-matching. To address this issue, we need to include additional conditions that further restrict the position of the node within an AST. We utilize both control-flow graph (CFG) and data-flow graph (DFG) information to enhance the rules.

Essentially, the CFG describes certain structured patterns that are loosely related, such as the statements within the \code{then} block if there is a change to the \code{if} conditional expression. The DFG describes the dependencies of data objects referenced outside of the modified sub-tree, for example, an assignment of a variable prior to its later usage or validation. We will delve into the details in Section~\ref{sec:cfgdfg}.

On the other hand, we expect the rule to be more generic so that it can accommodate variant instances with aliasing or other minor changes. Currently, we employ three basic forms of generalization mechanisms:

\begin{itemize} [leftmargin=*]
    \item Use equivalent or less strict QL built-in predicates. For example, for an equality expression \code{target} of \code{A==B}, the generalized condition is \code{`target.getAnOperand().getTarget()=A`} instead of \code{`target.getLeftOperand().getTarget()=A`}.
    \item Discard meaningless targets and conditions. Some code changes may not be directly relevant to the bug, but it is not always possible to preprocess them. In such cases, we discard the corresponding predicates. For instance, the predicate associated with updating \code{`!email->data`} to \code{`email->data == NULL`} in Listing~\ref{list:opensslpatch} is discarded.
    \item Substitute wrapper symbol names. Many projects use third-party or customized wrapper functions instead of general interfaces, which need to be replaced. For example, in Listing~\ref{list:opensslpatch}, the call to \code{OPENSSL\_strdup} is replaced with \code{strdup}.
\end{itemize}

\subsubsection{Regression tests}
Finally, we have a refined \qlname{} rule that is ready for testing. However, further refinement is still necessary such an elementary rule. Some predicates and query conditions may be redundant, and the CFG and DFG information for certain nodes may not be sufficient to narrow down the reported scope.

At this stage, we utilize regression tests to automatically refine the rule further. While ensuring that the rule can still match the target function, we try adding more CFG and DFG dependencies to see if false positives can be eliminated. We also combine potentially redundant predicates and conditions to check if any new false positives emerge. This iterative process is repeated until we obtain a reasonable rule in the end.

%

After the entire procedure, a set of viable rules for all known vulnerabilities was prepared. Later on, we can apply these rules to the \qltoolname{} databases of any project that is directly or indirectly related to the original OSS. For instance, by utilizing the rule to scan LibreSSL, we identified a seeming defect in its own copy of the very function \code{append\_ia5}~\cite{libresslvuln} (The concern was already reported to OpenBSD).

%% file: keyTechniques.tex
\section{Key Techniques}
\label{sec:keytech}

The implementation of \toolname{} took advantage of the EditScript algorithm with a customized matching method and a simplified method to retrieve CFG\&DFG information. In this section we go into details of these two key techniques.

\subsection{Customized matching method for EditScript algorithm}
\label{sec:editscript}

The original algorithm, called \textit{EditScript}, can be used to generate a correct edit script for hierarchically structured information, including SAST. However, correctness alone is not sufficient when it comes to source code comparison.

Let's revisit the example in Listing~\ref{list:opensslpatch}. The basic procedure of the EditScript algorithm follows a top-down, greedy approach, treating all nodes equally. When matching nodes in this manner, we encounter a discrepancy when it comes to the statement node \code{`return 1`} in Line 540 of the unpatched version. Instead of pairing it with its actual replication in Line 549 of the patched version, the algorithm pairs it with the inserted \code{`return 1`} node in Line 535. This logical mismatch is technically correct, but it causes us to overlook the fact that the entire \code{`if (memchr(...) != NULL return 1;`} block was inserted as a means to enforce data validation.

\begin{algorithm}[tb]
    \caption{Multi-stage method of matching 2 ASTs.}\label{algo:match}
    \footnotesize
    \begin{algorithmic}
    \State \textbf{Definitions:} 
    \State $D(Node)$ equals the number of all descendants of $Node$
    \State $Des_N(Node)$ is the $Nth$ descendant when DFS traversing the sub-tree rooted at $Node$
    \State $Node_a = Node_b$ if their types, values and number of children equal
    \State $Node_a \simeq Node_b$ if their types and number of children equal
    \State
    \Function{MatchUnit}{$Node_a,Node_b$}
        \If{$Node_a=Node_b$ and $Parent(Node_a)=Parent(Node_b)$ and for $N\leq D(Node_a)$ $Des_N(Node_a)=Des_N(Node_b)$}
        \State \Return $1$
        \ElsIf{$Node_a=Node_b$ and for $N\leq D(Node_a)$ $Des_N(Node_a)=Des_N(Node_b)$}
        \State \Return $2$
        \ElsIf{$Node_a\simeq Node_b$ and for $N\leq D(Node_a)$ $Des_N(Node_a)\simeq Des_N(Node_b)$}
        \State \Return $3$
        \ElsIf{$Node_a=Node_b$ and $Parent(Node_a) \simeq Parent(Node_b)$}
        \State \Return $4$
        \ElsIf{$Node_a=Node_b$}
        \State \Return $5$
        \ElsIf{$Node_a \simeq Node_b$}
        \State \Return $6$
        \EndIf
    \EndFunction
    \State
    \State $M \gets \emptyset$
    \For{$i$ in range$(1..3)$}
        \For{$Node_a$ in DFS traversing $Tree_a$}
            \If{$Node_a \notin M$ and $\exists Node_b \notin M$ and \Call{MatchUnit}{$Node_a,Node_b$}=$i$}
            \For{$N\leq D(Node_a)$} 
                \State $M \gets [Des_N(Node_a),Des_N(Node_b)]$
            \EndFor
            \EndIf
        \EndFor
    \EndFor
    \For{$i$ in range$(4..6)$}
        \For{$Node_a$ in DFS traversing $Tree_a$}
            \If{$Node_a \notin M$ and $\exists Node_b \notin M$ and \Call{MatchUnit}{$Node_a,Node_b$}=$i$}
            \State $M \gets [Node_a,Node_b]$
            \EndIf
        \EndFor
    \EndFor
    \end{algorithmic}
\end{algorithm}

To this end, we propose a customized matching method that allows for more accurate pairing of AST nodes. The objective in this step is to create a set of pairs, where each node from one AST is paired with the best matching node from the other AST. Any remaining unpaired nodes are considered singular nodes. This forms the foundation of the main procedure in the EditScript algorithm. The method is presented in Algorithm~\ref{algo:match}.

The underlying idea of the matching method is to consider additional context when conducting the search. If a subtree closely resembles another subtree, we can directly pair up the two nodes in the same position within those sub-trees, even if they may differ in values, types or predicates.

\subsection{Simplified CFG\&DFG analysis}
\label{sec:cfgdfg}

There are two primary reasons why contextual information is necessary to enhance the conditional query of a node. Firstly, it is used to establish correlations between predicates that query standalone nodes, enabling us to narrow down the target. This is particularly crucial in order to avoid false positives when querying a moved node or an inserted node that happens to replicate an existing one. Secondly, patched code is often placed upstream or downstream of the actual defects to block certain conditions or exploitation paths. These areas do not carry the most significant information about the defect. By tracing the control and data flow around the patched position, both backward and forward, we increase the chances of identifying the root cause of the defect.

While it is possible and convenient to retrieve CFG and DFG information using \qltoolname{}, we currently do not excessively complicate the task. The analysis is performed directly on the AST, using more intuitive methods.

\subsubsection{CFG analysis}
Currently, we provide additional information for AST nodes that belong to certain featured control structures. Specifically, the following control flow relations have been taken into consideration:

\begin{itemize} [leftmargin=*]
    \item If the targeted node is the conditional expression or its sub-expression of \code{If/Switch/For/While} statements, reference the featured statements within the \code{then/do} body. Conversely, if the targeted node resides within a \code{then/do} body, then reference its controlling conditional expressions.
    \item If the targeted node is the \code{lvalue} or \code{rvalue} of an assignment or comparison operation, then reference its sibling node.
    \item If the parent of the targeted node is an expression around an array, then reference its sibling node (array base or offset).
    \item More generally, for a node of \textit{expression} type, reference the outermost expression which wraps the target.
    \item In case more contextual information is required to further narrow down the target, for a node of \textit{statement} type, reference its direct parent and sibling statements, if any.
\end{itemize}

\subsubsection{DFG analysis}
Currently, we have not utilized precise DFG analysis. Instead, we simply reference the direct assignments and uses of key variables accessed in each block of differential nodes. The analysis is performed according to the type of access, as follows:

\begin{itemize} [leftmargin=*]
    \item If a variable is accessed as an \code{rvalue}, we trace back the control flow to find its nearest references as both \code{lvalue} and \code{rvalue} respectively. A LValue reference is likely an assignment operation, which determines the value/attributes of the variable used within the scope. A \code{rvalue} reference might be a validation operation. If the \code{lvalue} reference is closer, then the \code{rvalue} reference node is ignored.
    \item If a variable is accessed as an \code{rvalue}, we trace forward the control flow to find its nearest reference. If the reference turns out to be an \code{rvalue}, then it becomes a candidate for a DFG contextual node.
    \item If a variable is accessed as an \code{lvalue}, we trace forward the control flow to find its following \code{rvalue} references until a \code{lvalue} reference is encountered. We mark all the related nodes as candidates for supplementary DFG nodes.
\end{itemize}

%% file: evaluation.tex
\section{Evaluation and Observation}
\label{sec:evaluation}

In the following, we present the evaluation results of \toolname{}. The effectiveness can be evaluated based on the coverage and detection. Since the rule generation phase is conducted offline to the scanning phase, and the scanning is performed by invoking \qltoolname{}, performance is not a major consideration. What is more important is that by presenting the experimental results on extensively-tested projects, especially showcasing the newly-found defects, we can demonstrate the novelty of the tool and unveil the landscape of cognate defects in open-source software.

\input{eval1}

\input{eval2}

\input{eval3}

%% file: eval1.tex
\subsection{Accuracy and Recall}

We compare \toolname{} with three tools accessible for OSS: Coverity~\cite{coverity}, which is a SAST product that has been extensively validated in both commercial and open-source domains; VUDDY~\cite{kim2017vuddy}, which utilized a signature-based technique to carry out scalable code clone detection and had been referenced and compared extensively in research works in recent years; TRACER~\cite{kang2022tracer}, a research achievement that represents the research trend in the past two years, where vulnerability code features are designed to integrate information represented by taint analysis traces. The three tools represent the major approaches widely used to solve cognate defect problems: using general-purpose SAST tool and rules to discover vulnerabilities of certain patterns, using textual and structural fuzzy matching approaches to discover code with similarities, and using richer information of vulnerability features implied by control flow and data flow to discover semantically similar codes.

\smallskip \noindent
\textbf{Data set}\tab
To measure the benchmark accuracy of each target in detecting cognate defects, we selected 8 real-world open-source C language projects. These 8 projects have more than 30 historical CVE vulnerabilities each, of which vulnerability patches can be collected, and the project code has high complexity. Here we uniformly selected the first development version of each project after 2019, so it should be affected by a portion of the newly discovered vulnerabilities since 2020. As for the targeted defects to be matched, we collected all CVE vulnerabilities in year 2020 to June 2023, and only those which were fixed by applying patches on source code (but not Makefile or configurations) and the changed code were compiled with the default configuration. The total number of CVEs are indicated by $CVE\#$.

For each tool, we evaluated the total number of detection, i.e. positives. Among them we tried confirming those results accurately hitting the functions where CVE vulnerabilities resided within and got patched, i.e. true positives (TP). Specially, in the case of Coverity, which provide a significant number of defects without referring to any relevant CVE, we attempted to filter out those reports that hit vulnerable functions corresponding to a CVE vulnerabilities and while the reported cause complies with the vulnerability type.

\begin{table*}
    \caption{Accuracy in cognate defects detection of Coverity, VUDDY, Tracer and \toolname{}}\label{tab:comp} 
    \centering
    \begin{tabular}{c|c|c||c|c||c|c||c|c||c|c|c} \hline
          \multirow{2}{*}{Target} & \multirow{2}{*}{commit} & \multirow{2}{*}{CVE\#} &  \multicolumn{2}{|c||}{Coverity}&  \multicolumn{2}{|c||}{VUDDY}&  \multicolumn{2}{|c||}{TRACER}& \multicolumn{3}{|c}{Patch2QL}\\ \cline{4-12}
          &&&  TP&  Positives&  TP&  Positives&  TP&  Positives& TP& Positives&TP2\\ \hline \hline
          ImageMagick&dfd6399&  44&  0&  157&  8&  12&  0&  0& 7& 23&21\\ \hline
          curl&e24ea70&  40&  1&  120&  4&  10&  0&  4& 15& 17&34\\ \hline
          FFmpeg&7b58702&  52&  0& 749&  9&  13&  1&  1& 16& 23&35\\ \hline
          OpenSSL&12ad22d&  21&  0& 305&  4&  51&  1&  3& 4& 13&15\\ \hline
          ghostscript&d7d012b&  29&  0&  648&  6&  13&  \multicolumn{2}{|c||}{Error Running}& 12& 23&21\\ \hline
          libtiff&0a8245b&  28&  0&  45&  1&  4&  0&  6& 10& 22&23\\ \hline
          libxml2&8f62ac9&  11&  0&  215&  1&  1&  0&  32& 11& 15&11\\ \hline
          openjpeg&024b840&  9&  0&  25&  8&  16&  0&  0& 5& 9&7\\ \hline
    \end{tabular}
\end{table*}

The experimental data are listed in Table~\ref{tab:comp}, which we can interpret from three perspectives. Since whether the snapshot version of codebase was affected by a later CVE is hard to tell,\footnote{Because many vulnerabilities tended to affect only newly developed features, and there are hardly any methods or standardized data available for accurately determining the commits ranges affected by vulnerabilities.} a statistical false-negative rate may not have practical significance.

\smallskip \noindent
\textbf{Effective positives}\tab
From the table, we can see that compared to the other three benchmark tools, the rules generated by \toolname{} achieved the highest number of accurate detections for each target project, while maintaining the lowest total positives count. As a result, it achieved an evaluation accuracy rate of $90\%$. Because this benchmark experiment only focuses on the L1 and L2 types of code clones defined in the previous chapters, which ignore code differences caused by development evolution and target specific defective code in the ontology project, we can observe that similarity signature-based methods represented by VUDDY have achieved good results. On the other hand, semantic similarity-based methods represented by TRACER have difficulty leveraging their advantages in detecting fuzzy code clones.

We can also see that general-purpose SAST tools and rules, represented by Coverity, generated a large number of reports after performing comprehensive analysis on the target code. However, the proportion of reports that actually hit known vulnerabilities is relatively low. This reflects that in open-source projects that have been continuously analyzed by tools like Coverity, newly disclosed vulnerabilities are often not covered by generic vulnerability types and rules specific to programming languages. These vulnerabilities tend to be biased towards logical causes or involve causal analysis of complete vulnerability origins, which may exceed the capability boundaries of traditional SAST tools.

\smallskip \noindent
\textbf{Recall rate}\tab
To measure the recall rate of the \toolname{} rules, we conducted an additional experiment. We used the generated raw rules to scan the version immediately before each security vulnerability patch for each project (which should certainly be affected by the vulnerability). We then counted the cases where a successful match to the vulnerability (since 2020) could be made. These cases are separately defined as \textit{TP2} in the table.

A vulnerability is considered being \textit{recalled} only if the core vulnerable function(s) in the unpatched version are matched with its corresponding rule. Overall, the recall rate of rules generated by \toolname{}, evaluated by $TP2/CVE\#$, reaches $71.4\%$. A comprehensive evaluation on all $111$ projects yields a recall rate of $68.0\%$.

As indicated by experimental results in Table~\ref{tab:comp}, the recall rate of all CVE vulnerabilities ranges from $47\%$ (ImageMagick) to $100\%$. Such fluctuations in data are further discussed and explained in Section~\ref{sec:discussion}.

\smallskip \noindent
\textbf{\textit{False} positives}\tab
We used the rule set of all CVE vulnerabilities, not after 2020, to scan the target projects. The \textit{false} positives include 2 parts of detections: Those CVEs disclosed before 2020 matching the codebases, and those matching the code occurrences other than the very vulnerable functions.

The majority of these FP cases can be attributed to common root causes. In certain projects, temporary patches applied to a function may have been refactored to adhere to the coding style, and common fixing codes may have been abstracted into a separate function. In such cases, the \qlname{} could falsely identify the refactored function without sufficient interprocedural analysis.

Additionally, for certain vulnerabilities, their patches may not have provided enough information about the root causes or the very source position of the bug. These patches may have only implied the presence of a bug within the control flow. Some other patches may have aimed at blocking the exploitation path instead of directly fixing the bug. In these situations, the \qlname{} could incorrectly report other functions that have similar syntax to the vulnerable one, but require more contextual information to match more accurately.

Most of the auto-generated rules falling into the two categories mentioned above could be fine-tuned to eliminate the false positives while still identifying the target defect. However, there were some unintended results that could not easily be explained as \textit{false} positives. A portion of these results turned out to be actual cognate defects, which will be discussed further below.

%% file: eval2.tex
\subsection{New cognate defects found upstream}

\begin{table*}[ht]
    \caption{Resolved cognate defects.}
    \label{tab:cvelist}
    \centering
        \begin{threeparttable}
        \begin{tabular}{p{1.9cm}|p{2.6cm}|p{1cm}|p{2.6cm}|p{1cm}|p{6cm}}
            \hline
OSS & Identifier & Severity & Original CVE & Fix & Description \\ \hline \hline
ImageMagick & CVE-2023-34153 & High & CVE-2020-29599 & d31c80 & An incomplete sanitizing issue recurred in a newly added feature. \\ \hline
ImageMagick & CVE-2023-34152 & Critical & CVE-2020-29599 & 17c485 & Incomplete fix of CVE-2016-5118. \\ \hline
ImageMagick & CVE-2023-34151 & Medium & CVE-2022-32546 & 3d6d98 & Similar bad type-casting issues across decoders of formats. \\ \hline
opensc & CVE-2023-2977 & High & CVE-2021-42782 & 81944d & Buggy code reused in another module. \\ \hline
ghostscript & CVE-2023-38559 & Medium & CVE-2020-16305 & d81b82 & Similar OOB issues across devices. \\ \hline
ghostscript & CVE-2023-38560 & Medium & CVE-2017-9619 & b7eb1d & Patch for PCL handling bug not ported to XPS. \\ \hline
vim & CVE-2023-3896 & High & CVE-2023-0512 & 8154e6 & See below. \\ \hline
openssl & issue \#21111 & / & CVE-2015-1794 & 43596b & Similar divide-by-zero issues across crypto setup methods. \\ \hline
curl & issue \#11195 & / & CVE-2022-27780 & 6375a6 & Patch not ported to a redundant check point.\\ \hline
Linux Kernel & Pending & / & Various & Various & A growing list of Patches~\cite{kernelpatches}\\ \hline
\end{tabular}
\end{threeparttable}
\end{table*}

We have conducted an experiment by applying the rules of historical vulnerabilities against the latest development branch of codebase of respective projects. In a brief survey of the scanning results of all $111$ fundamental OSS, some occurrences were identified as new cognate defects, which are listed in the Table~\ref{tab:cvelist}. All of these defects have been promptly reported to the project maintainers and subsequently fixed. In certain cases, we were able to create proof-of-concept (PoC) examples to trigger the bugs, which were then classified as vulnerabilities and assigned CVE IDs. There were other bugs with clear root causes that were easily identifiable. However, demonstrating their exploitability would have required extensive expertise in the specific projects. Therefore, we reported them to the maintainers as \textit{potential} security bugs. Most of these cases were recognized as regular bugs and were appropriately addressed and fixed.

Table~\ref{tab:cvelist} presents the analyzed defects that were confirmed and reported as of June 30, within a span of approximately 30 days. We are still conducting analysis on additional potential defects. Due to the possibility of hackers discovering undisclosed vulnerabilities from the scan reports, we have temporarily suspended uploading the \qlname{} files that produced false positive results to the public repository.

From this list, we have made several general observations. Firstly, projects with extensive sub-modules or similar functions are more susceptible to cognate defects. When implementing a sub-module, such as a parser for a specific format or a device driver, developers often reuse existing code from similar modules. This reuse can occur at the file, function, block, and underlying logic levels. These cases belong to the L2--L3 type of cognate defects as defined in Section~\ref{subsec:threatmodel}, caused by copycat of following developers.

Secondly, it is not common practice within the open source community to conduct a code review specifically targeting cognate defects once a new vulnerability has been reported. Many instances of code cloning and derivation within projects lack documented records, making it challenging for maintainers to track such occurrences, especially when codes are copied from other authors.

Thirdly, a common lack of security consciousness has led to the emergence of cognate defects. For instance, consider CVE-2020-29599: In the parsing process of a PDF format, user input strings passed through an option are sanitized before being concatenated into a delegate command for execution. However, the sanitizer fails to check for illegal characters, such as the commonly used back-quote character for command injection. Approximately one year after the initial vulnerability was fixed, a new feature was added to support VIDEO formats. Unfortunately, the same mistake reoccurred in the new module, resulting in CVE-2023-34153. Similarly, the feature enabling pipes also sanitized user-supplied file names for later use as commands. However, the sanitizer once again missed checking for the presence of back-quote, leading to CVE-2023-34152. This case clearly confirms the scenario described in Section~\ref{subsec:threatmodel}, where cognate defects are caused by common weakness.

\begin{listing}[tb]
    \begin{minted}[frame=lines,
                       fontsize=\scriptsize, breaklines,
                       tabsize=2]{diff}
diff --git a/src/move.c b/src/move.c
index d3648df8b..3c50d258c 100644
--- a/src/move.c
+++ b/src/move.c
@@ -1933,6 +1933,9 @@ adjust_skipcol(void)
    return;

     int        width1 = curwin->w_width - curwin_col_off();
+    if (width1 <= 0)
+        return;  // no text will be displayed
+
     int     width2 = width1 + curwin_col_off2();
     long    so = get_scrolloff_value();
     int     scrolloff_cols = so == 0 ? 0 : width1 + (so - 1) * width2;
@@ -1976,5 +1979,5 @@ adjust_skipcol(void)
    if (col > width2)
    {
        row += col / width2;
        col = col % width2;
    }
    \end{minted}
    \caption{The patch of CVE-2023-0512 with extra contexts.} 
    \label{list:vimpatch}
    \vspace{-0.2em}
\end{listing}

\smallskip \noindent
\textbf{Case study}\tab
Let's examine the related defects discovered in VIM more closely. The initial vulnerability, CVE-2023-3896, was a divide-by-zero bug. This occurred when the cursor was moved within a VIM window, resulting in the recalculation of its width. The new width, which could be less than or equal to zero, was subsequently used as a divisor to calculate the number of rows. The patch that resolves this vulnerability can be found in Listing~\ref{list:vimpatch}.

Given that the root cause of the vulnerability is the absence of a check against zero for the variable \code{width1}, the main aspect of the rule is to identify a comparable declaration or assignment. The generated rule is appended in Listing~\ref{list:vimrule} in the appendix for reference.

\begin{listing}[t]
    \begin{minted}[frame=lines,
                    linenos=true, firstnumber=2592, numbersep=-10pt, breaklines,
                       fontsize=\scriptsize,
                       tabsize=2]{c++}
    int skip_lines = 0;
    int width1 = curwin->w_width - curwin_col_off();
    int width2 = width1 + curwin_col_off2();
    // similar formula is used in curs_columns()
    if (curwin->w_skipcol > width1)
      skip_lines += (curwin->w_skipcol - width1) / width2 + 1;
    \end{minted}
    \caption{CVE-2023-3896 in \code{scroll\_cursor\_bot}.}
    \label{list:vimbug}
\end{listing}

Through the use of the \qlname{} rule, we were able to identify a cognate defect in another function called \code{scroll\_cursor\_bot}. The vulnerable code snippet can be seen in Listing~\ref{list:vimbug}. The only similarity between this function and \code{adjust\_skipcol} is the calculation of updated \code{width1} and \code{width2}. There are no indications that the two functions share replicated code, making it extremely difficult to discover the vulnerability in the latter function using any code similarity-based or signature-based techniques. This discovery clearly demonstrates the effectiveness of \toolname{} in detecting L3--L4 levels of cognate defects which shared the weakest similarity.

%% file: eval3.tex
\subsection{Cognate defects found downstream}

By employing a set of rules for OSS, it becomes feasible to conduct a thorough investigation into the impact of vulnerabilities originating from upstream projects on downstream open-source ecosystems. To provide a clearer illustration of real-world implications and identify the most significant cases, we have selected two types of software widely utilized in cloud computing: databases and Linux kernels. These software systems are renowned for their extensive scale and complex codebases. Many vendors often fork established upstream projects and make substantial modifications in their own copies, which adds complexity to the task of confirming the existence or resolution of CVEs in such downstream projects.

\subsubsection{Databases} \

Regarding databases, we have selected two upstream OSS projects based on their common utilization downstream in distinct forms. The first is sqlite3~\cite{sqlite}, which is commonly reshaped and embedded into third-party programs in source code form. The second is PostgreSQL~\cite{postgresql}, which serves as the foundation for various commercial databases' extensions.

\begin{table}[tb]
    \caption{Defects found in projects derived from sqlite3 \& PostgreSQL.}
    \label{tab:forksdb}
    \centering
        \begin{threeparttable}
        \begin{tabular}{p{2cm}||p{1.1cm}|p{1.3cm}|p{1.4cm}}
            \hline
OSS & State & Upstream Version & Defects \\ \hline \hline
SQLCipher & c7f9a1 & 3.41.2 & 0 \\ \hline
libSQL & c734a2 & 3.43.0 & 0 \\ \hline
comdb2 & eb6e89 & 3.28.0 & 10 \\ \hline
Tarantool & 852664 & / & 7 \\ \hline
unqlite & 057067 & / & 1 \\ \hline
\hline
AgensGraph & 247f32 & 13.9 & 0 \\ \hline
Greenplum & b6acf7 & 12.12 & 0 \\ \hline
Kunlun & 5a4a4d & 11.5 & 15 \\ \hline
postgre-xl & 31dfe4 & 10.9 & 16 \\ \hline
TBase & 08b8c7 & 10.0 & 20 \\ \hline
PolarDB & c39c0d & 11.9 & 6 \\ \hline
openGauss & c61c75 & 9.2.4 & 3 \\ \hline
\end{tabular}
\end{threeparttable}
\end{table}

In a brief experiment conducted in Sep 2023, we gathered a collection of OSS projects that appear to be derived from sqlite3 or PostgreSQL. Only projects with a significant number of stars and active development were included in our selection. Since most derived projects do not explicitly maintain fork information, we primarily relied on featured strings for our search. The analysis results are presented in Table~\ref{tab:forksdb}.

A substantial portion of these downstream projects are hosted by corporate groups, and we observed a noteworthy difference in the response to bug and vulnerability reports compared to upstream developers. Although all defects have been reported to the project owners through GitHub issues, only a few have been acknowledged, which leaves them categorized as \textit{potential} defects.

Some projects have adopted good practices in synchronizing with upstream OSS. For instance, projects like SQLCipher, libSQL, AgensGraph, and Greenplum have consistently used relatively up-to-date versions of the codebase. They have cherry-picked or manually back-ported patches from upstream, even those that are only applicable to the latest branches. On the other hand, some projects exhibit common issues that make them not only susceptible to known vulnerabilities but also challenging to manually examine for the presence of CVEs. Several examples illustrating such common issues are listed below.

\smallskip \noindent
\textbf{Case study}\tab
Tarantool~\cite{tarantool} serves as a representative example of projects that customize and embed third-party open-source components in source code form. In this case, the sqlite3 project is utilized as a SQL parser. However, all copyright disclaimers have been replaced, and all instances of 'sqlite' strings within function and variable names have been removed. As a result, it becomes impossible to establish a clear correspondence between Tarantool and sqlite3.

\begin{listing}[tb]
    \begin{minted}[frame=lines, breaklines,
                       fontsize=\scriptsize,
                       tabsize=2]{diff}
diff --git a/src/select.c b/src/select.c
index 5f51074a0..44fb06f48 100644
--- a/src/select.c
+++ b/src/select.c
@@ -6065,5 +6065,6 @@ int sqlite3Select(
   if( (p->selFlags & (SF_Distinct|SF_Aggregate))==SF_Distinct
    && sqlite3ExprListCompare(sSort.pOrderBy, pEList, -1)==0
+   && p->pWin==0
   ){
     p->selFlags &= ~SF_Distinct;
     pGroupBy = p->pGroupBy = sqlite3ExprListDup(db, pEList, 0);
    \end{minted}
    \caption{The patch of CVE-2019-19244 in sqlite3.} 
    \label{list:sqlitepatch}
    \vspace{-0.2em}
\end{listing}

\begin{listing}[tb]
    \begin{minted}[frame=lines,
                    linenos=true, firstnumber=5833, numbersep=-10pt, breaklines,
                       fontsize=\scriptsize,
                       tabsize=2]{c++}
    if ((p->selFlags & (SF_Distinct | SF_Aggregate)) == SF_Distinct
        && sqlExprListCompare(sSort.pOrderBy, pEList, -1) == 0) {
        p->selFlags &= ~SF_Distinct;
        pGroupBy = sql_expr_list_dup(pEList, 0);
        p->pGroupBy = pGroupBy;
    \end{minted}
    \caption{The vulnerable function \code{sqlSelect} in tarantool.} 
    \label{list:tarantoolbug}
\end{listing}

The lack of mapping information presented a challenge for developers to identify the existence of upstream bugs. Despite active maintenance of the embedded code in Tarantool, some patches were still missing. For instance, the patch for CVE-2019-19244~\cite{sqlitecve} in the \code{sqlite3Select} function (Listing~\ref{list:sqlitepatch}) was found to be absent in the corresponding \code{sqlSelect} function in Tarantool (Listing~\ref{list:tarantoolbug}). The changes in the API names referenced caused code obfuscation, but the generated \qlname{} rules were still able to match them. This example can serve as a representative of L2-level cognate defects. It highlights the presence of residual defects after extensive code refactoring and secondary development in the project.

Another example is openGauss~\cite{opengauss}, an open-source RDBMS. Its kernel originated from an early version of PostgreSQL, which no longer receives upstream support. The PostgreSQL codebase was reconstructed in C++ and has received consistent maintenance from the community, including refined patches to accommodate the old codebase. However, instances of unfixed vulnerabilities can still be found despite the reconstruction efforts. 

\subsubsection{Kernels of Linux distributions}\label{subsec:kernel} \

The Linux kernel ecosystem constitutes a unique software landscape. In addition to the intricate branching upstream, each Linux distribution develops one or more kernels of its own, resulting in a significant level of kernel fragmentation. When it comes to maintaining stable packages of open-source software, including kernels, most distributions such as RHEL~\cite{rhel} and SLES~\cite{sles}, choose to diverge directly from the upstream kernel and backport security patches. However, all these distributed kernels adhere to a specific patch version of the upstream kernel and do not rebase to newer versions, as doing so would entail the effort of merging changes. Consequently, the responsibility of backporting upstream patches falls upon community developers.

\begin{table}[tb]
    \caption{Potential kernel defects found in Linux distributions of cloud environments.}\label{tab:kerneldists}
    \centering
        \begin{tabular}{p{2.0cm}||p{1.2cm}|p{0.9cm}|p{0.5cm}|p{0.5cm}|p{0.5cm}}
            \hline
\multirow{2}*{OS} & \multirow{2}*{Commit} & \multirow{2}*{\thead{Derived\\From}} & \multicolumn{3}{c}{Defects} \\ 
\cline{4-6} 
&&& \tiny $\geq$High & \tiny Medium & \tiny Low \\
\hline \hline
openEuler & 03ad78 & 5.10.0 & 0 & 5 & 0 \\ \hline
OpenAnolis & 89caf8 & 4.19.91 & 12 & 12 & 1 \\ \hline
OpenCloudOS & 320119 & 5.4.119 & 12 & 14 & 1 \\ \hline
\end{tabular}
\end{table}

At the end of Feb 2023, we conducted an experiment using three open-source Linux kernels released by communities hosted by the three largest cloud computing vendors in China. We utilized a selected set of \qlname{} rules that covered CVE vulnerabilities and other security bugs~\cite{osvdev} in the upstream Linux kernel starting from 2021. The results are presented in Table~\ref{tab:kerneldists}.

After analyzing the results and examining the git histories of each distribution, we drew three conclusions. First, the vulnerability policy adopted by each distribution is a significant factor in determining its level of security. Among the three distributions, openEuler~\cite{openeuler} consistently merged security patches from the upstream \code{5.10.y} branch. In contrast, OpenAnolis~\cite{openanolis} and OpenCloudOS~\cite{opencloud} preferred to merge a batch of patches in one go before releasing a new version. Second, most defects were discovered in outdated codebases, including original code that had been updated upstream and backported, but left unmaintained features from higher versions. The customized part of the kernels primarily involved adding drivers for third-party devices and subsystems, rather than rewriting mainline code. This made it relatively easier to validate the existence of defects as long as the dependent kernel versions remained within the upstream life cycle. Third, the extensive version fragmentation led to numerous complexities. If a feature was backported from a later version, maintainers often lost track of upstream security patches. Furthermore, if maintainers created patches for zero-day vulnerabilities ahead of the Linux community upstream, conflicts could arise during subsequent code synchronization, potentially leading to the reoccurrence of old vulnerabilities.

\smallskip \noindent
\textbf{Case study}\tab
One reported and confirmed defect in openEuler is CVE-2022-1015~\cite{cve20221015} (referred to as openEuler-SA-2023-1253~\cite{openeulerbug}). The original CVE was fixed on March 17, 2022, and initially disclosed on April 29, 2022. According to the report, this bug only affected mainline kernels starting from v5.12-rc1, and the patch was applied to 5.15, 5.16, and 5.17 LTS kernels. At that time, it was believed that the kernel in openEuler, based on the 5.10 upstream version, was unaffected. However, the results indicated that the vulnerability pattern perfectly matched the code. Subsequently, on April 18, 2023, openEuler maintainers acknowledged the report and merged the mainline fix~\cite{fix20221015}. It was not until June 28, 2023, that the upstream community cherry-picked the fix to the maintenance branch linux-5.10.y~\cite{fix20221015up}, resulting in a gap of 468 days.

\begin{listing}[tb]
    \begin{minted}[frame=lines, breaklines,
                       fontsize=\scriptsize,
                       tabsize=2]{diff}
diff --git a/arch/x86/kernel/cpu/resctrl/rdtgroup.c b/arch/x86/kernel/cpu/resctrl/rdtgroup.c
index 064e9ef44cd6..9d4e73a9b5a9 100644
--- a/arch/x86/kernel/cpu/resctrl/rdtgroup.c
+++ b/arch/x86/kernel/cpu/resctrl/rdtgroup.c
@@ -3072,7 +3072,8 @@ static int rdtgroup_rmdir(struct kernfs_node *kn)
    * If the rdtgroup is a mon group and parent directory
    * is a valid "mon_groups" directory, remove the mon group.
    */
-   if (rdtgrp->type == RDTCTRL_GROUP && parent_kn == rdtgroup_default.kn) {
+   if (rdtgrp->type == RDTCTRL_GROUP && parent_kn == rdtgroup_default.kn &&
+       rdtgrp != &rdtgroup_default) {
        if (rdtgrp->mode == RDT_MODE_PSEUDO_LOCKSETUP ||
            rdtgrp->mode == RDT_MODE_PSEUDO_LOCKED) {
                ret = rdtgroup_ctrl_remove(kn, rdtgrp);
    \end{minted}
    \caption{The patch of the bug in Intel RDT.} 
    \label{list:rdtpatch}
    \vspace{-0.2em}
\end{listing}

\begin{listing}[tb]
    \begin{minted}[frame=lines, breaklines,
                       fontsize=\scriptsize,
                       tabsize=2]{c++}
static int resctrl_group_rmdir_ctrl(struct kernfs_node *kn, struct resctrl_group *rdtgrp,
                   cpumask_var_t tmpmask)
{
    resctrl_group_rm_ctrl(rdtgrp, tmpmask);
    kernfs_remove(rdtgrp->kn);
    return 0;
}

static int resctrl_group_rmdir(struct kernfs_node *kn)
{
    ...
    if (rdtgrp->type == RDTCTRL_GROUP && parent_kn == resctrl_group_default.kn)
        ret = resctrl_group_rmdir_ctrl(kn, rdtgrp, tmpmask);
    else if (rdtgrp->type == RDTMON_GROUP &&
         is_mon_groups(parent_kn, kn->name))
        ret = resctrl_group_rmdir_mon(kn, rdtgrp, tmpmask);
    ...
}
    \end{minted}
    \caption{The cognate snippet in the borrowed \code{resctrlfs}.} 
    \label{list:openeulerbug}
\end{listing}

Another noteworthy case is the identification of a questionable cognate snippet. In addition to CVE vulnerabilities, we also considered security bugs identified in GSD~\cite{gsd} for rule generation. In openEuler, developers implemented a user interface for ARM v8 MPAM in the \textit{fs/resctrlfs.c} file. The implementation was partially borrowed from the upstream kernel's implementation of Intel RDT, as announced. We discovered a rule regarding a bug that existed in the latter, as shown in Listing~\ref{list:rdtpatch}, and we matched a cognate snippet in the former, as shown in Listing~\ref{list:openeulerbug}. After an assessment, the community maintainers concluded that this piece of code does not constitute a vulnerability. However, it still serves as an example to demonstrate the possibility of cognate vulnerabilities arising in the real world when code is reused.

%% file: discussion.tex
\section{Discussion}
\label{sec:discussion}

\smallskip \noindent
\textbf{Recall rate}\tab
As shown in Table~\ref{tab:comp}, the recall rates varied significantly across different projects. This variation can be attributed to the average complexity of both the codebases and the patches. In general, patches that are either overly simple or overly complex tend to have lower quality.

For overly simple patches, where only a single statement or expression is inserted, there may not be enough features or related CFG / DFG context for the rule to be translated into meaningful predicates and query conditions. To measure the richness of information carried within patches, we calculated the average number of lines changed (including added and deleted lines) for all target vulnerability patches in the eight experimental projects. For ImageMagick, the average number of lines changed per patch is $40.5$, and the average number of lines of code (LoC) changed per related function is $27.9$, while the results for the libtiff are $143.7$ and $46.2$ respectively. This implies that there can be a large portion of patches in ImageMagick changing so little code that the generated rules got dropped for producing many false positives.

On the other hand, overly sophisticated patches may indicate complex changes or even a complete rewrite of the code logic. These patches could involve too many elements to be adequately represented within the generated rule. In some other cases, developers of specific projects opt to push a single huge commit where vulnerability fixes and other irrelevant changes got mixed up. In the extreme case of PostgreSQL, it is calculated that every patch involves $8.2$ functions and $935.0$ LoC. These all resulted in the difficulty in separating core code changes out of the patch and make a rule.

\smallskip \noindent
\textbf{Overhead}\tab
In the evaluation section, we did not specifically evaluate or compare the execution efficiency of the tools. This is primarily because the scanning process is implemented by CodeQL, which we rely on, and generating rules based on patches using \toolname{} is done offline so the overhead can be ignored. However, considering that signature-based fuzzy matching algorithms and SAST rule matching are both computationally intensive mechanisms, their execution efficiency is still worth discussing.

In the experiments, we set CodeQL to use 8 parallel threads in analyzing each target project, and more than $98\%$ of the rules finished execution within 30 minutes. This is actually faster than average built-in rules of CodeQL, since data-flow analysis is only used in the generation but not the execution phase of the rules. The scan results of Coverity are typically returned to the user within half an hour after uploading the intermediate data of the project and starting the scan. As for custom signature-based tools, the situations are quite different. For VUDDY, the local signature caculation phase can be done in 10 minutes on average, and the analysis result is returned to user instantly after uploading. For TRACER, however, time and memory consumption can be undesirable. On a server with 80 CPU cores and 350GB memory installed, one analysis task took up half of the cores and over $60\%$ memory at most, and in an extreme case the analysis of ghostscript could not be finished.

From the above comparison, it can be observed that with signature-based fuzzy matching methods, the computational overhead may become uncontrollable as the signatures carry more information and the project size increases. On the other hand, SAST methods based on syntactic and semantic pattern matching have better expectations, and the rules themselves in SAST are more readable. We also expect a much lower overhead if CodeQL is replaced by some specific tools like Clang AST Matcher~\cite{astmatcher}.

\smallskip \noindent
\textbf{Scalability}\tab
As explained in Section~\ref{sec:burden}, the OSS issues in the C/C++ language have been chosen as our primary research target, mainly due to the historical burden of the corresponding software supply chain. However, our approach is not limited to a specific programming language.

From a technical perspective, the underlying tool we currently use, CodeQL, provides analysis capabilities for all mainstream language ecosystems, including AST dump and analysis. \toolname{} itself utilizes a cross-language, generic approach for structured AST comparison and QL query combination. Additional adaptation and optimization are required for generating language-specific syntax rules, such as how to perform forward and backward traversing control flow, which we believe will only require a certain amount of development effort.

From an objectives perspective, our preliminary analysis targeting Java and other ecosystems has revealed that cognate defects are also prevalent, although their forms may vary. For example, in Java, there are cases where explicitly introduced packages coexist with their older versions implicitly introduced in source-code form, leaving hidden threats. These issues can also be explored using the existing capabilities of \toolname{}.

%% file: conclusions.tex
\section{Conclusion}
\label{sec:conclusions}

In this paper, we uncover the origins and risks of cognate defects in OSS. We propose a comprehensive procedure for transforming vulnerability patches into rules for \qltoolname{}, capturing key semantic and dependent contextual information. We have developed a prototype tool called \toolname{}. Evaluating the application of our system on popular OSS written in C has demonstrated its effectiveness and unique insights. It has achieved a recall rate of $68\%$ for historical CVEs and better effectiveness than general-purpose SAST rules and signature-based mechanisms. It identified $7$ new vulnerabilities, and uncovered additional issues in upstream projects, downstream databases, and Linux kernels, including various cognate defects.


\section*{Availability}

The rules for historical vulnerabilities of OSS (including the Linux kernel) generated by \toolname{} have been open sourced in the repository~\cite{qlrulesrepo} and are available for open source developers to use freely.

%% file: appendix.tex
\section*{Appendix}

\begin{appendices}

\begin{listing}[h]
    \begin{minted}[frame=lines,
                    breaklines,
                    breakafter=.(=,
                    fontsize=\scriptsize,
                    tabsize=2]{sql}
import cpp

predicate func_0(Variable vwidth1_1935, AddExpr target_1) {
  exists(IfStmt target_0 |
    target_0.getCondition().(RelationalOperation).getLesserOperand().(VariableAccess).getTarget()=vwidth1_1935
    and target_0.getCondition().(RelationalOperation).getGreaterOperand().(Literal).getValue()="0"
    and target_0.getCondition().(RelationalOperation).getLesserOperand().(VariableAccess).getLocation().isBefore(target_1.getAnOperand().(VariableAccess).getLocation()))}

predicate func_1(Variable vwidth1_1935, AddExpr target_1) {
    target_1.getAnOperand().(VariableAccess).getTarget()=vwidth1_1935
    and target_1.getAnOperand().(FunctionCall).getTarget().hasName("curwin_col_off2")}

predicate func_2(Variable vwidth1_1935, Initializer target_2) {
    vwidth1_1935.getInitializer()=target_2
    and target_2.getExpr().(SubExpr).getLeftOperand().(PointerFieldAccess).getName()="w_width"
    and target_2.getExpr().(SubExpr).getRightOperand().(FunctionCall).getTarget().hasName("curwin_col_off")}

from Function func, Variable vwidth1_1935, AddExpr target_1, Initializer target_2
where
not func_0(vwidth1_1935, target_1)
and func_1(vwidth1_1935, target_1)
and func_2(vwidth1_1935, target_2)
and vwidth1_1935.getType().hasName("int")
and vwidth1_1935.(LocalVariable).getFunction() = func
select func
    \end{minted}
    \caption{The generated \qlname{} rule matching CVE-2023-0512.} 
    \label{list:vimrule}
    \vspace{-0.2em}
\end{listing}

\begin{figure}[h]
    \centering
    \includegraphics[height=0.95\linewidth,angle=90]{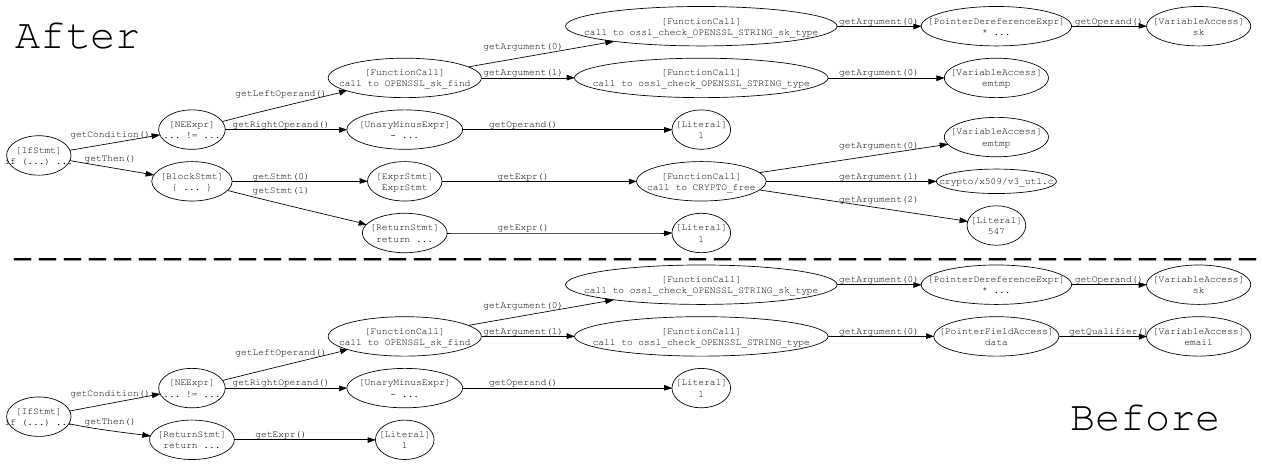}
    \caption{A pair of ASTs of pre- and post-patched functions}
    \label{fig:asts}
\end{figure}

\end{appendices}